\def\year{2021}\relax
\documentclass[letterpaper]{article} 
\usepackage{aaai21}  
\usepackage{times}  
\usepackage{helvet} 
\usepackage{courier}  
\usepackage[hyphens]{url}  
\usepackage{graphicx} 
\usepackage{natbib}  
\usepackage{caption} 
\usepackage{booktabs}
\usepackage{algorithm}
\usepackage{bbding}
\usepackage{graphicx}
\usepackage{subcaption}  
\usepackage{multirow}
\usepackage{amsmath,amsthm}

\title{GraphCrop: Subgraph Cropping for Graph Classification}

\author{
	Yiwei Wang\textsuperscript{\rm 1}, Wei Wang\textsuperscript{\rm 1}, Yuxuan Liang\textsuperscript{\rm 1}, Yujun Cai\textsuperscript{\rm 2}, Bryan Hooi\textsuperscript{\rm 1}
	\\
}
\affiliations{
	\textsuperscript{\rm 1}School of Computing, National University of Singapore, Singapore\\
	\{y-wang,wangwei,yuxliang,bhooi\}@comp.nus.edu.sg\\
	\textsuperscript{\rm 2}Nanyang Technological University, Singapore\\
	yujun001@e.ntu.edu.sg
	
}

\begin{document}
	
\maketitle

\begin{abstract}
We present a new method to regularize graph neural networks (GNNs) for better generalization in graph classification.
Observing that the omission of sub-structures does not necessarily change the class label of the whole graph, we develop the \textbf{GraphCrop} (Subgraph Cropping) data augmentation method to simulate the real-world noise of sub-structure omission.
In principle, GraphCrop utilizes a node-centric strategy to crop a contiguous subgraph from the original graph while maintaining its connectivity.
By preserving the valid structure contexts for graph classification, we encourage GNNs to understand the content of graph structures in a global sense, rather than rely on a few key nodes or edges, which may not always be present.
GraphCrop is parameter learning free and easy to implement within existing GNN-based graph classifiers.
Qualitatively, GraphCrop expands the existing training set by generating novel and informative augmented graphs, which retain the original graph labels in most cases.
Quantitatively, GraphCrop yields significant and consistent gains on multiple standard datasets, and thus enhances the popular GNNs to outperform the baseline methods.
\end{abstract}

\section{Introduction}
\textbf{Graph classification} is a fundamental task on graph data, which aims to predict the class label of an entire graph.
The modern tools of choice for this task are graph neural networks (GNNs).
Typically, GNNs build node representations from node features and the graph topology via the `message passing' mechanism and then make graph-level predictions by summarizing the node representations through a readout function \cite{xu2018powerful}, \cite{sun2019infograph}.

GNNs are capable of making predictions based on complex graph structures, thanks to their advanced representational power.
However, the increased representational capacity comes with higher model complexity, which can induce over-fitting and weaken the generalization ability of GNNs.
In this case, a trained GNN may capture random error or noise 
instead of the underlying data distribution \cite{zhang2016understanding}, which is not what we expect.

\begin{figure}[!tb]
	\centering
	\includegraphics[width=\linewidth]{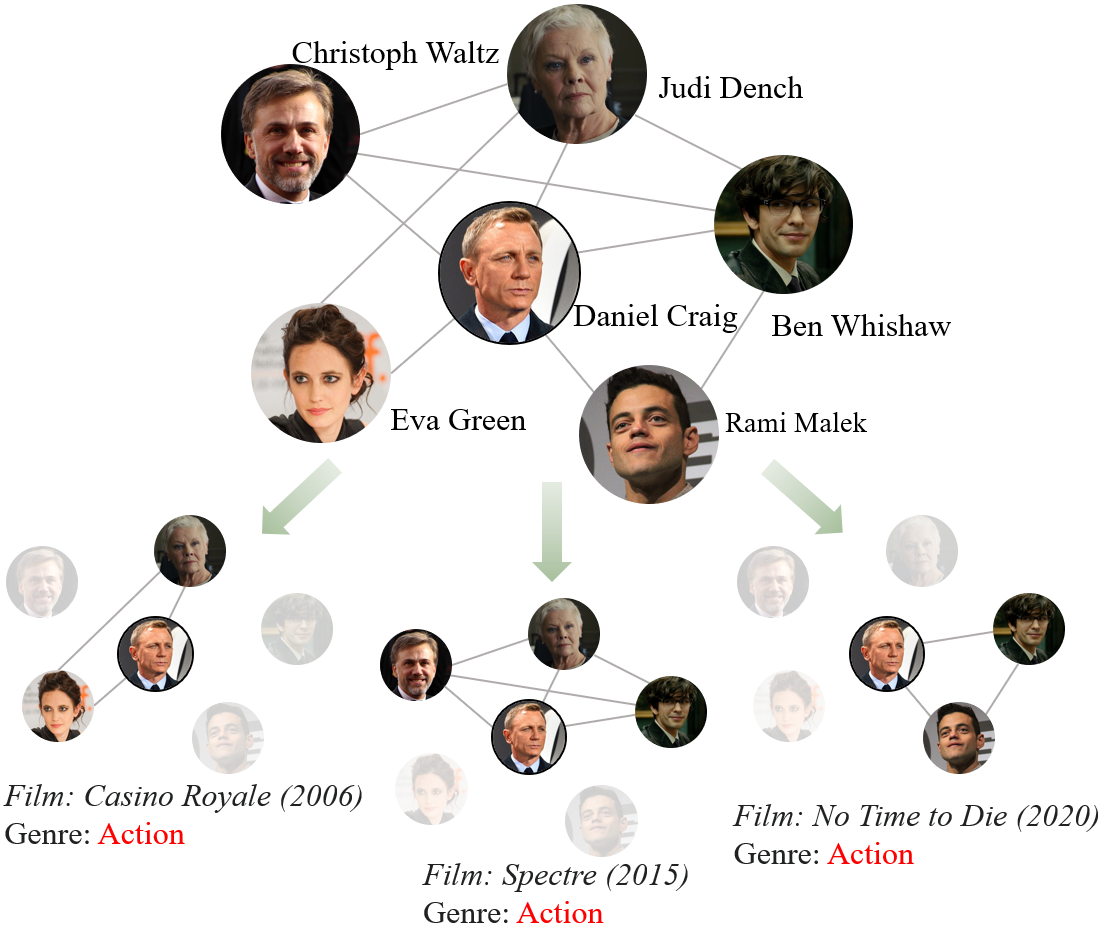}
	\caption{\textbf{Omission of sub-structures does not change the genre label `Action' of an actor Daniel Craig's ego-network \cite{yanardag2015deep}.} In the real-world movie collaboration datasets \cite{yanardag2015deep}, nodes represent actors/actresses, and there is an edge between them if they appear in the same movie. Class labels of graphs are the genres of the movies in which the actors/actresses collaborate.  \label{fig:ego}}
\end{figure}

To combat the over-fitting of neural networks, data augmentation has been demonstrated to be effective in the classification of image data \cite{perez2017effectiveness}.
Nevertheless, data augmentation for graph classification remains under-explored, of which a major challenge remains in the high irregularity of graph data.
On the one hand, effective data augmentation aims to make significant and diverse modifications to the features while keeping their ground-truth labels unchanged \cite{xie2019unsupervised}.
On the other hand, how to manipulate a graph without altering its class label is not clear.

In our work, we observe that the omission of partial nodes and edges (sub-structures), does not necessarily change the class label of the whole graph (see Fig. \ref{fig:ego}).
In a scientific collaboration dataset (\texttt{COLLAB}), for example, graphs are ego-networks of researchers, and class labels are the research fields  \cite{yanardag2015deep}.
Omission of partial researchers does not necessarily change the research field of the whole graph.
In addition, in terms of the real-world noise on graph data, missing partial sub-structures is likely to happen.
For instance, to collect an ego-network for a researcher, it is difficult to collect all of his/her collaborators without omission.
A satisfactory GNN is expected to classify graphs correctly even in the presence of this type of noise.
On the contrary, the collected training graphs may not exhibit enough variance in the absence of sub-structures.
As a result, the trained GNN is likely to classify the clean graphs correctly, but fail to recognize the graphs of incomplete structures due to its limited generalization ability.

\begin{figure}[!tb]
	\centering
	\includegraphics[width=\linewidth]{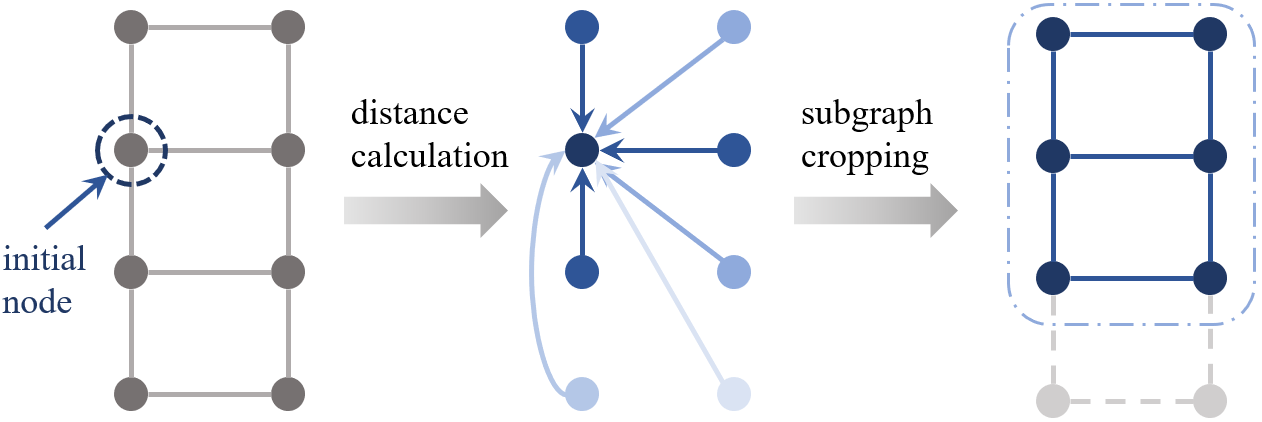}
	\caption{\textbf{Pipeline of GraphCrop.} In the middle column, the connectivity from different nodes to the initial node (in the darkest color) is positively related to their color shades. \label{fig:pip}}
\end{figure}

To address the above issues, we propose a novel data augmentation approach for graph classification, called \textbf{GraphCrop} (Subgraph Cropping), that uses a node-centric strategy to crop a contiguous
subgraph from the original graph.
GraphCrop can be easily implemented within existing GNNs to achieve better inference performance on graph classification for virtually free.
In the training phase, GraphCrop augments a graph within a mini-batch with probability $p$ and keeps it unchanged with probability $1-p$.
In the former case, we first select an \textit{initial node} randomly from the original graph, and then retain 
the nodes holding the \textit{strongest connectivity} to the initial node and the edges between them to form the cropped subgraph (see Fig. \ref{fig:pip}).
The number of nodes to retain in the subgraph grows as the size of the original graph, since larger graphs can maintain their original labels while absorbing more noise.
To evaluate the connectivity between nodes, we use the diffusion matrix as the metric, which is related to the random walk process over graphs \cite{pons2005computing}.
In this manner, the subgraphs spanning contiguous parts of the original graph
are cropped, resulting in the novel and informative samples to expand the existing training set.

Previous work includes DropEdge, a data augmentation technique for the task of node-level predictions, which drops each edge randomly at uniform distribution \cite{rong2019dropedge}.
Analogous to it, we refer to the method that drops nodes uniformly as UniNode.
Different from both of them, GraphCrop crops a contiguous subgraph which preserves the intact graph structure
rather than the individual nodes or edges as DropEdge and UniNode do.
GraphCrop ensures the connectivity within the subgraph and conveys valid structural contexts for graph classification.
In this fashion, GraphCrop encourages GNNs to understand the content of the graph structure in a global sense, rather than focusing on a few key nodes or edges, which may not always be present.
Otherwise, if the nodes or edges are removed uniformly at random, the original structural features may no longer be maintained with the broken topological characteristics.
We demonstrate the advantages of GraphCrop empirically.

We evaluate GraphCrop on the graph classification task using  the standard chemical \cite{dobson2003distinguishing} and social \cite{yanardag2015deep} datasets.
Qualitatively, GraphCrop generates novel augmented graphs, which simulate the real-world noise of sub-structure omission, and preserves the original class labels in most cases (see Fig. \ref{fig:tsne}).
Quantitatively, we observe the improvements in test accuracy of graph classification for GNN models trained with GraphCrop.
The improvements are higher than those given by the data augmentation designed for other tasks \cite{rong2019dropedge}.
Overall, GraphCrop improves the performance of the popular GIN \cite{xu2018powerful} and EigenPool \cite{ma2019graph} by a significant margin, and enhances them to outperform the baseline methods.

\section{Related Work}
\textbf{Graph Classification.} 
Early solutions to graph classification include graph kernels.
The pioneering work \cite{haussler1999convolution} decomposes graphs into small subgraphs and computes kernel functions based on their pair-wise similarities.
Subsequent work proposes various subgraphs, such as paths \cite{borgwardt2005shortest}, and subtrees \cite{shervashidze2009efficient}, \cite{neumann2016propagation}.
We refer readers to \cite{nikolentzos2019graph}, \cite{kriege2020survey} for a general overview.
More recently, many efforts have been made to design graph neural networks (GNNs) for graph classification \cite{scarselli2008graph}, \cite{li2015gated}, \cite{niepert2016learning}, \cite{gilmer2017neural}, \cite{ying2018hierarchical}, \cite{zhang2018end}, \cite{xu2018powerful}.
These work focuses on developing GNN architectures of higher complexity to improve their fitting capacity.
In contrast, our framework is orthogonal to them in the sense that we propose a new data augmentation method that enhances a GNN model by generating novel and informative augmented graphs for training.
As far as we know, we are the first to develop a data augmentation approach for graph classification.

\noindent\textbf{Data Augmentation.}
Data augmentation plays a central role in training neural networks. 
It only operates on the input data, without changing the model architecture, but improves the performance significantly. 
For example, in image classification,  strategies such as horizontal flips, random erasing \cite{zhong2017random}, Hide-and-Seek \cite{singh2018hide}, and Cutout \cite{devries2017improved} have been shown to improve performance. 
On graph data, \cite{wang2020nodeaug} proposes NodeAug to augment and utilize the unlabeled nodes in the task of semi-supervised node classification.
DropEdge \cite{rong2019dropedge} removes edges uniformly as data augmentation for node classification.
For graph classification, DropEdge is not effective since it tends to break the graph structural properties.
Different from it, we crop the contiguous subgraphs from the original graph to generate novel and valid augmented graphs, which simulate the real-world noise of sub-structure omission.

\begin{figure*}[!tb]
	\centering
	\includegraphics[width=\linewidth]{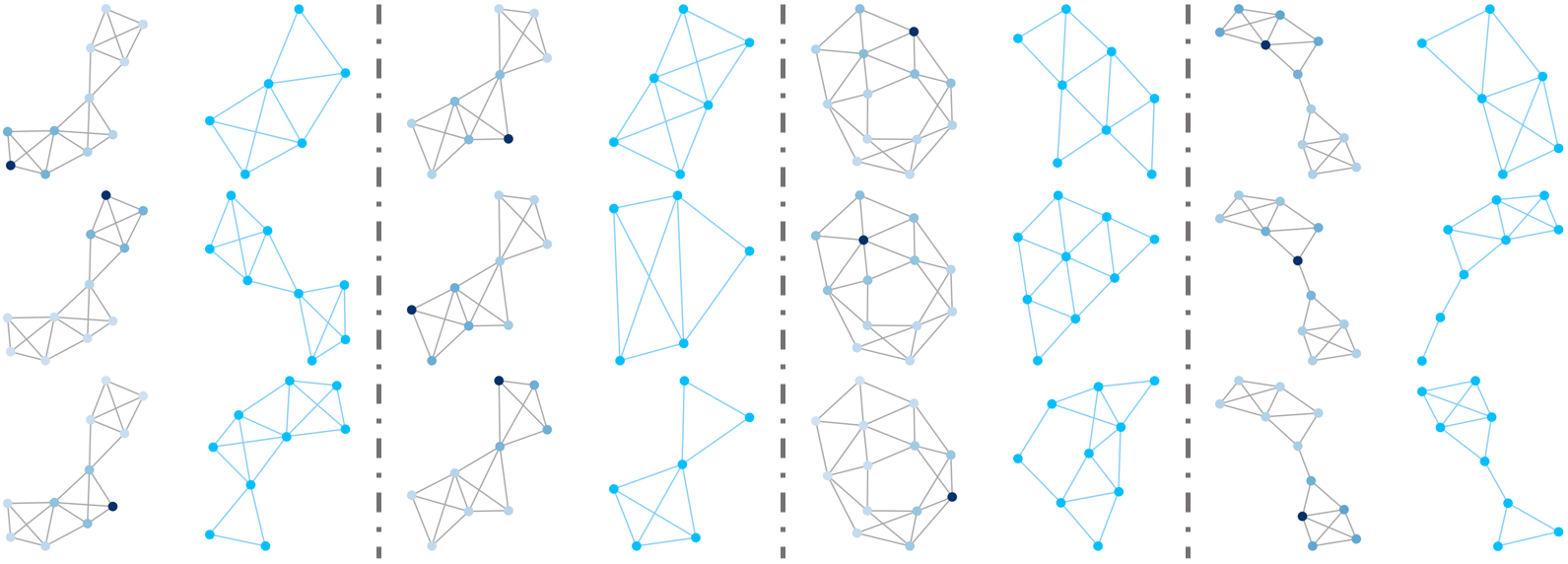}
	\caption{\textbf{Example subgraphs that GraphCrop obtains from the \texttt{PROTEINS} dataset.} The original graphs are in the odd-numbered columns, 
	while the corresponding cropped subgraphs are on their right hand side. The connectivity from different nodes to the initially selected node (in the darkest color) is positively related to their color shades. \label{fig:exa}}
\end{figure*}

\section{Methodology}
In this section, we describe the details of our GraphCrop (Subgraph Cropping) data augmentation method.
At each training step, GraphCrop augments a graph $G$ in the mini-batch with probability $p$, while keeping $G$ unchanged with probability $1-p$.
In this process, graph neural networks (GNNs) are exposed to novel and miscellaneous graphs randomly cropped from the original ones.

\subsection{Background and Motivation}
A graph $G$ consists of a set of nodes $\mathcal{V}$ and a set of edges $\mathcal{E}$.
Graph classification aims to learn a mapping function that maps every graph to a predicted class label.
Graph neural networks are the state-of-the-art solution for graph classification thanks to their advanced representational capacity. 
However, the increased representational power also comes with a higher probability of over-fitting, leading to poor generalization.
In order to combat the potential for over-fitting, we propose a data augmentation method named GraphCrop to regularize GNNs for better inference performance.

GraphCrop crops a subgraph $\hat{G}$ from $G$ and feeds $\hat{G}$ to GNNs for training.
The main motivation of GraphCrop is that the omission of partial nodes and edges (sub-structure) does not necessarily alter the class label of $G$.
We visualize an example in Fig. \ref{fig:ego} from the movie collaboration datasets, such as \texttt{IMDB-B}, where the graphs are ego-networks of actors/actresses and the class labels are the genre of the movies in which the actors/actresses collaborate  \cite{yanardag2015deep}.
The subgraphs of the original ego-network correspond to the real-world movies belonging to the ground truth genre `Action', inferring that the graph label does not change given the omission of some sub-structures.
We have similar observations in other kinds of graphs, such as the scientific collaboration datasets (\texttt{COLLAB}) and online discussion community datasets (\texttt{REDDIT-B}) \cite{yanardag2015deep}.
In the former one, for example, graphs are researchers' ego-networks and labels are their research fields.
Omission of partial researchers does not necessarily alter the research fields of the whole graph.
Besides, in chemical datasets, prior research has observed that chemical graphs in the same class share common subgraphs \cite{prvzulj2007biological}, implying the existence of unnecessary sub-structures for categorizing graphs.

In terms of the real-world noise on graph data, missing a partial sub-structure of a graph is likely to happen. 
For instance, to collect an ego-network for a researcher, it is difficult to collect all of his/her collaborators without omission.
For humans, this omission can be negligible because it does not break the connectivity and the topological characteristics of the whole graph.
And for GNNs, it is desirable that the models classify graphs correctly invariant to this kind of noise. 
However, the collected training graphs may exhibit limited variance in the omission of sub-structures.
As a result, the learned GNN is likely to work well on the graphs without omission, but fail to recognize the graphs of incomplete structures.
We aim to simulate the absence of sub-structures in practice to expand the existing training set.
In order to achieve it, we need to maintain connectivity and topological characteristics while conducting subgraph cropping.

\begin{figure*}[!tb]
	\centering
	\includegraphics[width=\linewidth]{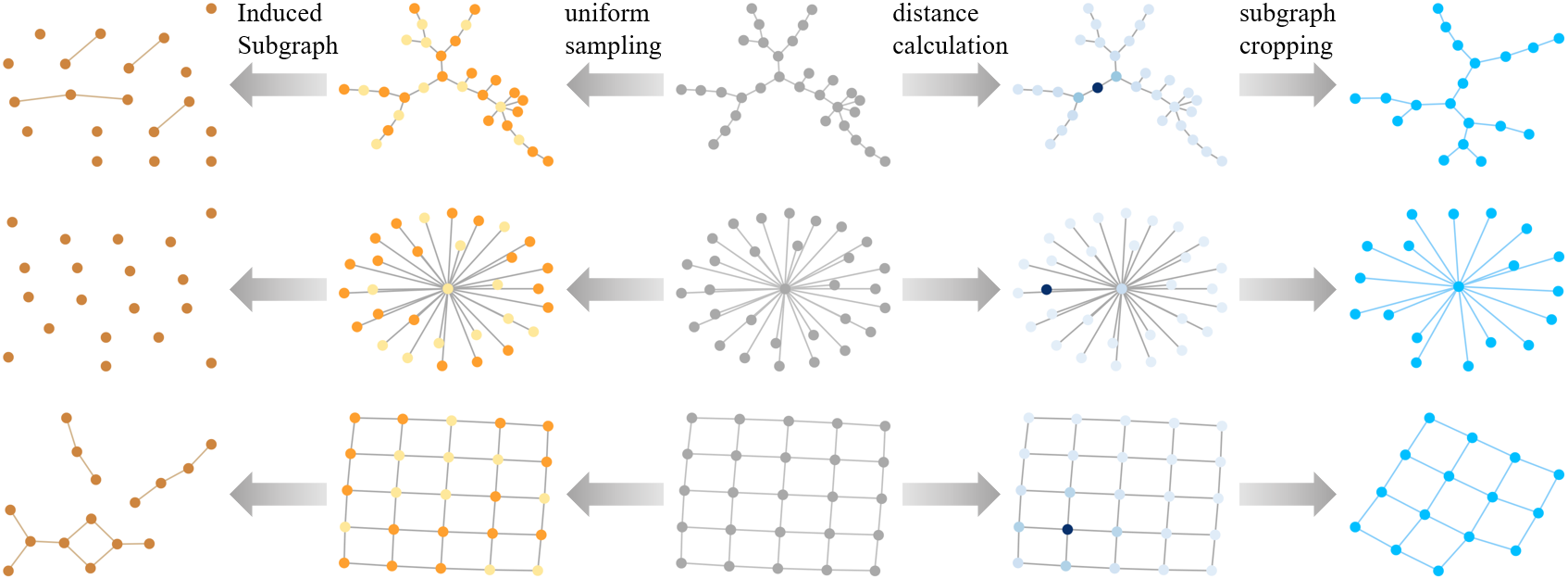}
	\caption{\textbf{GraphCrop retrieves the subgraphs (fifth column) maintaining the topology characteristics of original graphs (third column).} On the other hand, the subgraphs given by UniNode (first column) do not have the original structural property. In column 2, the sampled nodes are in darker colors. In column 4, the nodes' color shades are positively related to their connectivity to the initially selected node (in the darkest color). \label{fig:uni}}
\end{figure*}

\subsection{Subgraph Cropping}

When cropping a subgraph from the original graph $G$, we need to decide what nodes and edges to retain.
We design our GraphCrop method as the following steps: 

\noindent1. We select a node $v$ (the initial one) uniformly at random from the set of all nodes in $\mathcal{V}$;

\noindent2. We evaluate the connectivity from the nodes in $\mathcal{V}_i$ to $v$;

\noindent3. We select the $\lceil\rho \#\mathcal{V}\rceil$ nodes in $\mathcal{V}$ with the highest connectivity to $v$ and all the edges between them to form the subgraph, where $\rho \in [0,1]$ is a hyper-parameter to adjust the size of the cropped subgraph.

We visualize this pipeline in Fig. \ref{fig:pip}.
Since large graphs have more nodes than small ones, they can absorb more noise while maintaining their ground-truth label. 
To compensate, we vary the number of nodes in the subgraph by $\lceil\rho \#\mathcal{V}\rceil$.
To fulfill step (2), we need to use a metric to evaluate the connectivity between nodes.
A simple metric is the length of the shortest paths \cite{johnson1973note}.
However, this metric does not necessarily change when connections (paths) between nodes increase, and thus it does not reflect the connectivity effectively.
In our work, we utilize the diffusion matrix to measure the connectivity \cite{fouss2012experimental}:
\begin{equation}
\mathbf{S} = \sum_{k = 0}^{\infty}\Theta_k \mathbf{T}^k,
\end{equation}
where $\mathbf{T}$ is the generalized transition matrix and $\Theta_k$ is the weighting coefficient which determines the ratio of global-local information.
Personalized PageRank (PPR) \cite{page1999pagerank} and heat kernels \cite{kondor2002diffusion} are two popular instantiations of graph diffusion.
Given an adjacency matrix $\mathbf{A}$, and a diagonal degree matrix $\mathbf{D}$, PPR and heat diffusion are defined by setting $\mathbf{A}\mathbf{D}^{-1}$, and $\theta_k = \alpha(1-\alpha)^k$ and $\theta_k = e^{-t}$ respectively, where $\alpha$ denotes teleport probability in a random work and $t$ is diffusion time \cite{klicpera2019diffusion}.
The closed-form solution to the PPR diffusion is formulated as \cite{hassani2020contrastive}:
\begin{equation}
\mathbf{S}^{\text{PPR}} = \alpha(\mathbf{I} - (1 - \alpha)\mathbf{D}^{-1/2}\mathbf{A}\mathbf{D}^{-1/2})^{-1},\\
\end{equation}
where $\mathbf{I}$ is the identity matrix.

Graph diffusion is related to the random walk process \cite{pons2005computing}.
With $\mathbf{T} = \mathbf{A}\mathbf{D}^{-1}$, $T^{k}_{ij}$ is the probability that a random walk goes from $j$ to $i$ in $k$ steps.
A large $k$ achieves a sufficiently long random walk to gather enough information about the topology of the graph.
$\mathbf{S}$, as the weighted sum of $\mathbf{T}^{k}$ of different $k$, captures both the local and global connectivity information between nodes.
Prior research \cite{klicpera2019diffusion} validates that the graph learning can benefit from the connectivity retrieved by graph diffusion.

GraphCrop can be easily performed on the CPU during data loading.
By implementing this operation on the CPU in parallel with the main GPU training task, we can hide its computation and obtain performance improvements virtually for free.
Moreover, graph diffusion can be computed once during the pre-processing stage.
Existing techniques such as fast approximation and sparsification methods can improve its efficiency \cite{klicpera2019diffusion}.
We visualize examples of cropped subgraphs in Fig. \ref{fig:exa}.

\subsection{Discussion}
To augment the graph data, previous research proposes DropEdge \cite{rong2019dropedge} for node classification, which drops edges uniformly without removing nodes. 
Similar to it, we name the method to uniformly remove the nodes as UniNode, as visualized in Fig. \ref{fig:uni}.
The main distinction between GraphCrop and the above two methods is that we retain contiguous subgraphs rather than individual nodes, as demonstrated in Fig. \ref{fig:exa} and \ref{fig:uni}.
With GraphCrop, the nodes in the cropped subgraphs are connected strongly and the intact topology in the subgraphs is maintained.
This is motivated by the discovery from prior research that the subgraphs of strong connectivity contain rich features for recognizing the full graph \cite{zeng2019accurate}.
Hence, the subgraphs convey the intact and valid structural context for graph classification.
In this way, GraphCrop forces GNNs to understand the content of the graph structures in a global sense, rather than focusing on a few key nodes, which may not always be present.
Otherwise, if the nodes are selected randomly at uniform distribution, as shown in Fig. \ref{fig:uni}, the topology characteristics of the original graph is broken, and the original structural features are not maintained.

GraphCrop endows every node and edge with non-negligible probability to be sampled, which guarantees that no information is lost through our random data augmentation during training.
Overall, GraphCrop simulates the real-world sub-structure omission in the real-world graphs, i.e., it generates graphs that are novel and informative to GNNs.
From this vantage, we prepare GNNs for encounters with real-world noise during training, so as to enable GNNs to take more of the graph structural context into consideration when making decisions.

\section{Experiments}\label{sec:exp}

\begin{table}[tb!]
	\centering
	\caption{\textit{Statistics of the utilized datasets.} $\#$Nodes denotes the average number of nodes per graph, while $\#$Edges denotes the average number of edges per graph.}
	\label{tab:dataset}
	\begin{tabular}{@{}l c c c@{}}
		\toprule
		\textbf{Dataset}
		& \textbf{$\#$Graphs}
		& \textbf{$\#$Nodes}
		& \textbf{$\#$Edges}\\ \midrule\midrule
		\texttt{D$\&$D} & 1,178 & 284.32 & 715.66 \\
		\texttt{ENZYMES} & 600 & 32.63  & 62.14 \\
		\texttt{NCI1} & 4,110 & 29.87 & 32.30 \\
		\texttt{NCI109} & 4,127 & 29.68 & 32.13\\
		\texttt{PROTEINS} & 1,113 & 39.06 & 72.82\\
		\midrule
		\texttt{COLLAB} & 5,000 & 74.49 & 2457.78\\
		\texttt{IMDB-B} & 1,000 & 19.77 & 96.53\\
		\texttt{IMDB-M} & 1,500 & 13.00 & 65.94\\
		\texttt{REDDIT-B} & 2,000 & 429.63 & 497.75\\
		\texttt{REDDIT-5K} & 4,999 & 508.52 & 594.87\\
		\bottomrule
	\end{tabular}
\end{table}

\begin{table*}[tb!]
	\centering
	\caption{\textit{Test Accuracy (\%) of graph classification on chemical datasets.} We perform 10-fold cross-validation to evaluate model performance, and report the mean and standard derivations over 10 folds. We highlight best performances in bold.}
	\label{tab:bio}
	\begin{tabular*}{\textwidth}{@{\extracolsep{\fill}} l c c c c c @{}}
		\toprule
		\textbf{Method}
		& \texttt{D$\&$D}
		& \texttt{ENZYMES}
		& \texttt{NCI1}
		& \texttt{NCI109}
		& \texttt{PROTEINS}
		\\ \midrule\midrule
		GRAPHLET \cite{shervashidze2009efficient} & 72.1 $\pm$ 3.7 & 41.4 $\pm$ 5.2 & 64.3 $\pm$ 2.2 & 62.5 $\pm$ 2.8 & 70.1 $\pm$ 4.1 \\ 
		WL \cite{shervashidze2011weisfeiler} & 73.2 $\pm$ 1.8 & 53.7 $\pm$ 6.0 & 76.3 $\pm$ 1.9 & 75.8 $\pm$ 2.3& 72.3 $\pm$ 3.4 \\
		GCN \cite{kipf2016semi} & 74.2 $\pm$ 3.1 & 59.1 $\pm$ 4.7 & 76.8 $\pm$ 2.1 & 76.1 $\pm$ 2.2 & 73.3 $\pm$ 3.6 \\
		DGCNN \cite{zhang2018end} & 76.7 $\pm$ 4.1 &39.3 $\pm$ 5.9& 76.5 $\pm$ 1.9 & 75.9 $\pm$ 1.7 & 72.9 $\pm$ 3.5 \\
		DiffPool \cite{ying2018hierarchical} & 75.2 $\pm$ 3.8 &59.7 $\pm$ 5.3& 76.8 $\pm$ 2.0 & 75.5 $\pm$ 1.9 & 73.6 $\pm$ 3.6\\
		EigenPool \cite{ma2019graph}  & 75.9 $\pm$ 3.9 & 62.4 $\pm$ 3.8 & 78.7 $\pm$ 1.9 & 77.4 $\pm$ 2.5 & 74.1 $\pm$ 3.1 \\
		GIN \cite{xu2018powerful} & 75.4 $\pm$ 2.6 &60.3 $\pm$ 4.2 & 79.7 $\pm$ 1.8 & 78.2 $\pm$ 2.1 & 73.5 $\pm$ 3.8 \\
		\midrule
		GraphCrop + GCN & 75.3 $\pm$ 3.0 & 60.1 $\pm$ 4.2 & 77.3 $\pm$ 2.1 & 77.2 $\pm$ 2.2 & 74.0 $\pm$ 3.5\\
		GraphCrop + EigenPool & \textbf{77.3 $\pm$ 3.2} & \textbf{63.6 $\pm$ 3.6} & 79.6 $\pm$ 2.0 & 78.5 $\pm$ 2.3 & \textbf{74.9 $\pm$ 3.3} \\
		GraphCrop + GIN & 76.6 $\pm$ 2.7 & 60.9 $\pm$ 3.1 & \textbf{80.9 $\pm$ 1.9} & \textbf{79.5 $\pm$ 2.2} & 74.1 $\pm$ 3.5\\
		\bottomrule
	\end{tabular*}
\end{table*}

\begin{table*}[tb!]
	\centering
	\caption{\textit{Test Accuracy (\%) of graph classification on social datasets.} We perform 10-fold cross-validation to evaluate model performance, and report the mean and standard derivations over 10 folds. We highlight best performances in bold.}
	\label{tab:soc}
	\begin{tabular*}{\textwidth}{@{\extracolsep{\fill}} l c c c c c @{}}
		\toprule
		\textbf{Method}
		& \texttt{COLLAB}
		& \texttt{IMDB-B}
		& \texttt{IMDB-M}
		& \texttt{REDDIT-B}
		& \texttt{REDDIT-5K}
		\\ \midrule\midrule
		GRAPHLET \cite{shervashidze2009efficient} & 61.7 $\pm$ 2.2 & 54.8 $\pm$ 4.1 & 42.6 $\pm$ 2.7& 62.1 $\pm$ 1.6 & 36.2 $\pm$ 1.8\\ 
		WL \cite{shervashidze2011weisfeiler} & 70.4 $\pm$ 1.8& 69.1 $\pm$ 3.5 & 45.4 $\pm$ 2.9& 81.7 $\pm$ 1.7 & 49.4 $\pm$ 2.1\\
		GCN \cite{kipf2016semi} & 74.3 $\pm$ 2.0 & 70.3 $\pm$ 3.7 & 48.2 $\pm$ 3.1 & 86.6 $\pm$ 1.9 & 53.7 $\pm$ 1.7\\
		DGCNN \cite{zhang2018end} &71.1 $\pm$ 1.7& 69.2 $\pm$ 2.8 & 45.6 $\pm$ 3.4 & 87.6 $\pm$ 2.1 & 49.8 $\pm$ 1.9 \\
		DiffPool \cite{ying2018hierarchical} & 68.9 $\pm$ 2.2& 68.6 $\pm$ 3.1& 45.7 $\pm$ 3.4& 89.2 $\pm$ 1.8& 53.6 $\pm$ 1.4 \\
		EigenPool \cite{ma2019graph} & 70.8 $\pm$ 1.9 & 70.4 $\pm$ 3.3& 47.2 $\pm$ 3.0 & 89.9 $\pm$ 1.9 & 54.5 $\pm$ 1.7 \\
		GIN \cite{xu2018powerful} &75.5 $\pm$ 2.3 & 71.2 $\pm$ 3.9& 48.5 $\pm$ 3.3& 89.8 $\pm$ 1.9 & 56.1 $\pm$ 1.6 \\
		\midrule
		GraphCrop + GCN & 75.1 $\pm$ 2.2 & 71.3 $\pm$ 3.5 & 48.8 $\pm$ 3.2 & 87.1 $\pm$ 1.8 & 54.6 $\pm$ 1.8\\
		GraphCrop + EigenPool & 71.7 $\pm$ 2.1 & 71.0 $\pm$ 3.1 & 47.9 $\pm$ 3.1 & 90.8 $\pm$ 1.8 & 55.2 $\pm$ 1.8\\
		GraphCrop + GIN & \textbf{76.9 $\pm$ 2.2} & \textbf{72.9 $\pm$ 3.7} & \textbf{49.6 $\pm$ 3.2} & \textbf{91.0 $\pm$ 1.9} & \textbf{57.7 $\pm$ 1.7}\\
		\bottomrule
	\end{tabular*}
\end{table*}

In this section, we first present the graph classification performance of GNN models trained with GraphCrop.
Next, we adjust the amount of labeled data to evaluate the generalization of GNNs with and without GraphCrop.
Then, we visualize the final-layer graph representations retrieved by GNNs and those augmented by GraphCrop.
Finally, we conduct ablation studies to show the influence of different node or edge samplers, as well as the sensitivity of the performance with respect to the hyper-parameters of GraphCrop.

We use the standard benchmark datasets: \texttt{D$\&$D} \cite{dobson2003distinguishing}, \texttt{ENZYMES} \cite{schomburg2004brenda}, \texttt{NCI1}, \texttt{NCI109} \cite{wale2008comparison}, \texttt{PROTEINS} \cite{borgwardt2005protein}, \texttt{COLLAB}, \texttt{IMDB-B}, \texttt{IMDB-M}, \texttt{REDDIT-B}, and \texttt{REDDIT-5K} \cite{yanardag2015deep} for evaluation.
The former five are chemical datasets, where the nodes have categorical input features.
The latter five are social datasets that do not have node features.
We follow \cite{xu2018powerful}, \cite{zhang2018end} to use node degrees as features.
The statistics of these datasets are summarized in Table \ref{tab:dataset}.

We use popular graph classification models as the baselines:
GRAPHLET \cite{shervashidze2009efficient} and Weisfeiler-Lehman Kernel (WL) are classical graph kernel methods, while GCN \cite{kipf2016semi}, DGCNN \cite{zhang2018end}, DiffPool \cite{ying2018hierarchical}, EigenPool \cite{ma2019graph}, and GIN \cite{xu2018powerful} are the GNNs with state-of-the-art performance in graph classification.
In addition, we include the recently proposed data augmentation method designed for node classification, DropEdge \cite{rong2019dropedge}, for comparison.

We use the PPR diffusion to measure connectivity by default.
For the hyper-parameters of graph diffusion and baselines, e.g., the number of layers, the optimizer, the learning rate, we set them as suggested by their authors.
For the hyper-parameters of our CurGraph, we set the augmentation probability $p = 0.5$, and $\rho = 0.7$ for the sizes of cropped subgraphs by default.

\begin{table*}[!h]
	\centering
	\caption{Results of node classification averaged over the 20 random splits of the varied ratio $r$ of labeled examples, in terms of test accuracy (\%). We highlight the best performance in bold.}
	\label{tab:random}
	\begin{tabular}{@{}l|cc|cc|cc@{}}
		\toprule 
		\multirow{2}{*}{\textbf{Method}} & \multicolumn{2}{c|} {\texttt{NCI1}} & \multicolumn{2}{c|} {\texttt{PROTEINS}} & \multicolumn{2}{c}
		{\texttt{COLLAB}} \\
		& $r = 60\%$ & $r = 80\%$ & $r = 60\%$ & $r = 80\%$ & $r = 60\%$ & $r = 80\%$ \\
		\midrule\midrule
		GCN \cite{kipf2016semi} & 68.4 $\pm$ 2.4 & 72.9 $\pm$ 2.2 & 65.8 $\pm$ 4.0 & 70.1 $\pm$ 3.9 & 67.7 $\pm$ 2.3 & 71.2 $\pm$ 2.2\\
		GraphCrop + GCN & 71.6 $\pm$ 2.2 & 74.3 $\pm$ 2.1 & 69.0 $\pm$ 3.8 & \textbf{71.4 $\pm$ 3.7} & 70.2 $\pm$ 2.3 & 72.5 $\pm$ 2.3\\
		\midrule
		GIN \cite{xu2018powerful} & 71.1 $\pm$ 2.2 & 75.5 $\pm$ 2.0 & 65.2 $\pm$ 4.1 & 69.8 $\pm$ 3.7 & 68.0 $\pm$ 2.5 & 71.8 $\pm$ 2.3 \\
		GraphCrop + GIN & \textbf{74.5 $\pm$ 2.0} & \textbf{77.1 $\pm$ 1.9} & \textbf{69.1 $\pm$ 3.8} & 71.2 $\pm$ 3.6 & \textbf{70.6 $\pm$ 2.4} & \textbf{73.7 $\pm$ 2.2} \\
		\bottomrule
	\end{tabular}
\end{table*}

\subsection{Graph Classification}
We follow \cite{xu2018powerful}, \cite{errica2019fair} to use the 10-fold cross-validation scheme to calculate the classification performance for a fair comparison.
For each training fold, as suggested by \cite{errica2019fair}, we conduct an inner holdout technique with a $90\%/10\%$ training/validation split, i.e., we train fifty times on a training fold holding out a random fraction ($10\%$) of the data to perform early stopping.
These fifty separate trials are needed to smooth the effect of unfavorable random weight initialization on test performances.
The final test fold score is obtained as the mean of these fifty runs.

We report the average and standard deviation of test accuracy across the 10 folds within the cross-validation on the chemical and social datasets in Table \ref{tab:bio} and \ref{tab:soc} respectively.
On the chemical datasets, we observe that GraphCrop improves the test accuracy of EigenPool by 1.8\% on \texttt{D\&D}, 1.9\% on \texttt{ENZYMES}, 1.1\% on \texttt{NCI1}, 1.4\% on \texttt{NCI109}, 1.1\% on \texttt{PROTEINS} respectively.
In addition, GraphCrop enhances GIN by 1.6\% on \texttt{D\&D}, 1.0\% on \texttt{ENZYMES}, 1.5\% on \texttt{NCI1}, 1.7\% on \texttt{NCI109}, 0.8\% on \texttt{PROTEINS}, and improves GCN by 1.5\% on \texttt{D\&D}, 1.7\% on \texttt{ENZYMES}, 0.7\% on \texttt{NCI1}, 1.4\% on \texttt{NCI109}, 1.0\% on \texttt{PROTEINS}.
On the social datasets, GraphCrop improves GCN by more than 1\% on \texttt{COLLAB}, \texttt{IMDB-B}, \texttt{IMDB-M}, and \texttt{REDDIT-5K}.
Meanwhile, GraphCrop achieves substantial improvements for EigenPool and GIN on all the social datasets.
As a result, CurGraph enhances EigenPool and GIN to outperform all the baseline methods.

Taking a closer look, we observe that the graph kernel methods, GRAPHLET and WL, generally present worse performance than the GNN methods.
This demonstrates the stronger fitting capacity of the advanced neural network models.
GraphCrop generally achieves higher improvements on EigenPool and GIN than that on GCN.
The reason is that EigenPool and GIN are the more advanced GNN models proposed for graph classification than GCN.
However, their increased learning power comes with higher risks of over-fitting.
Our GraphCrop effectively regularizes them by expanding the training set with the novel and informative augmented graphs, which reduces their over-fitting tendencies successfully.

\subsection{Training Set Sizing}

Over-fitting tends to be more severe when training on smaller datasets. 
By conducting experiments using a restricted fraction of the available training data, we show that GraphCrop has more significant improvements for smaller training sets.
We randomly select $r\in\{60\%, 80\%\}$ graphs from the whole set to form the labeled data, with the left graphs being the test graphs, to form a split.
For each labeled set, as suggested by \cite{errica2019fair}, we conduct an inner holdout technique with a $90\%/10\%$ training/validation split.
In other words, we train fifty times on a labeled set holding out a random fraction ($10\%$) of the data to perform early stopping.
These fifty separate trials are needed to smooth the effect of unfavorable random weight initialization on test performances.
We conduct the experiments on 20 random splits and report the mean and standard derivations over all splits in Table \ref{tab:random}.
Empirically, GraphCrop enhances the performance of GCN and GIN for different sizes of the training set. 
In principle, with less labeled nodes, i.e., smaller $r$, our method gives larger accuracy improvements, which demonstrates the necessity of the regularization given by our GraphCrop especially when the labeled data is limited.

\subsection{Visualization of GraphCrop}
\begin{figure}[!tb]
	\centering
	\begin{subfigure}[t]{0.234\textwidth}
		\includegraphics[width=\textwidth]{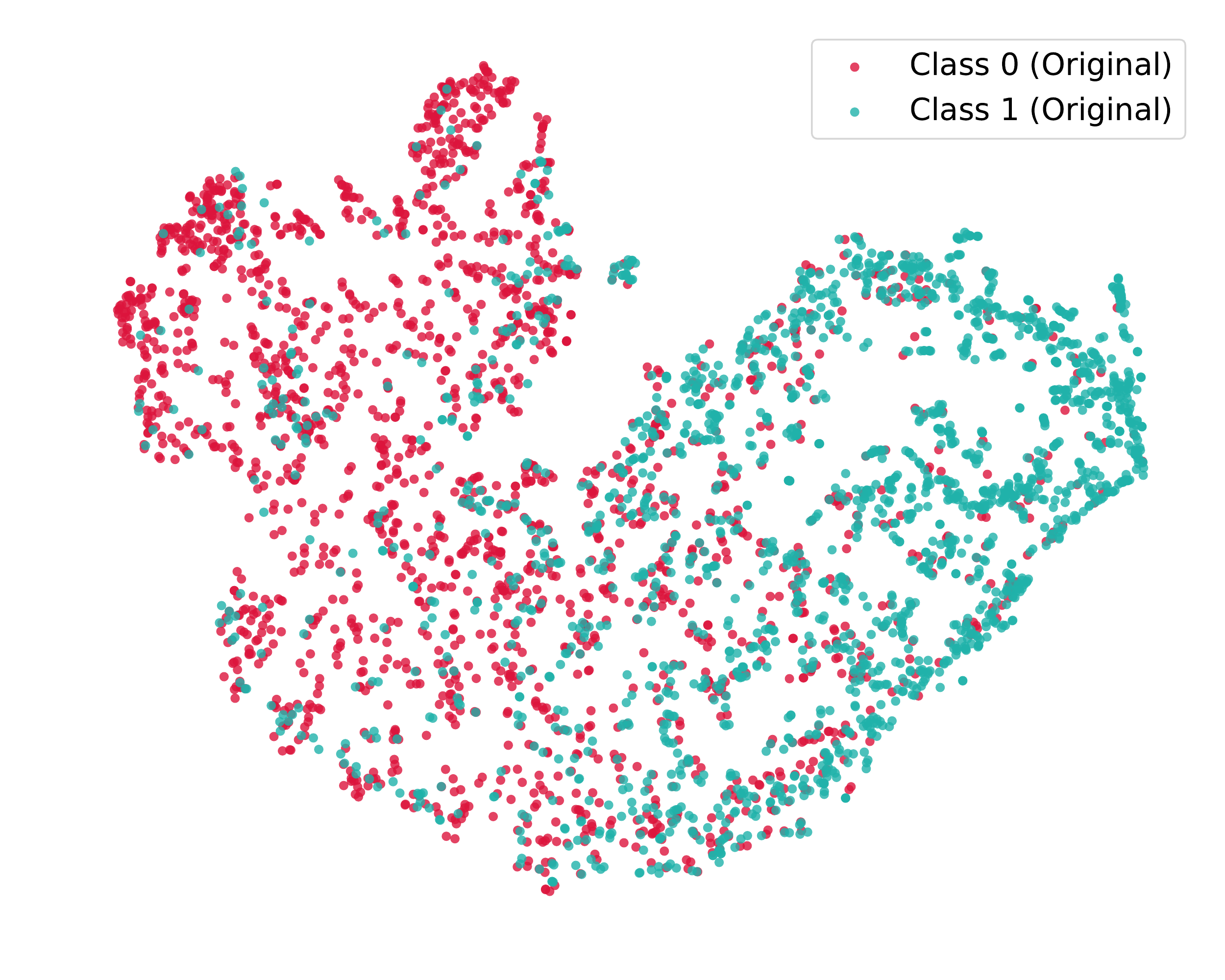}
		\caption{GIN}
	\end{subfigure}	\hfill
	\begin{subfigure}[t]{0.234\textwidth}
		\includegraphics[width=\textwidth]{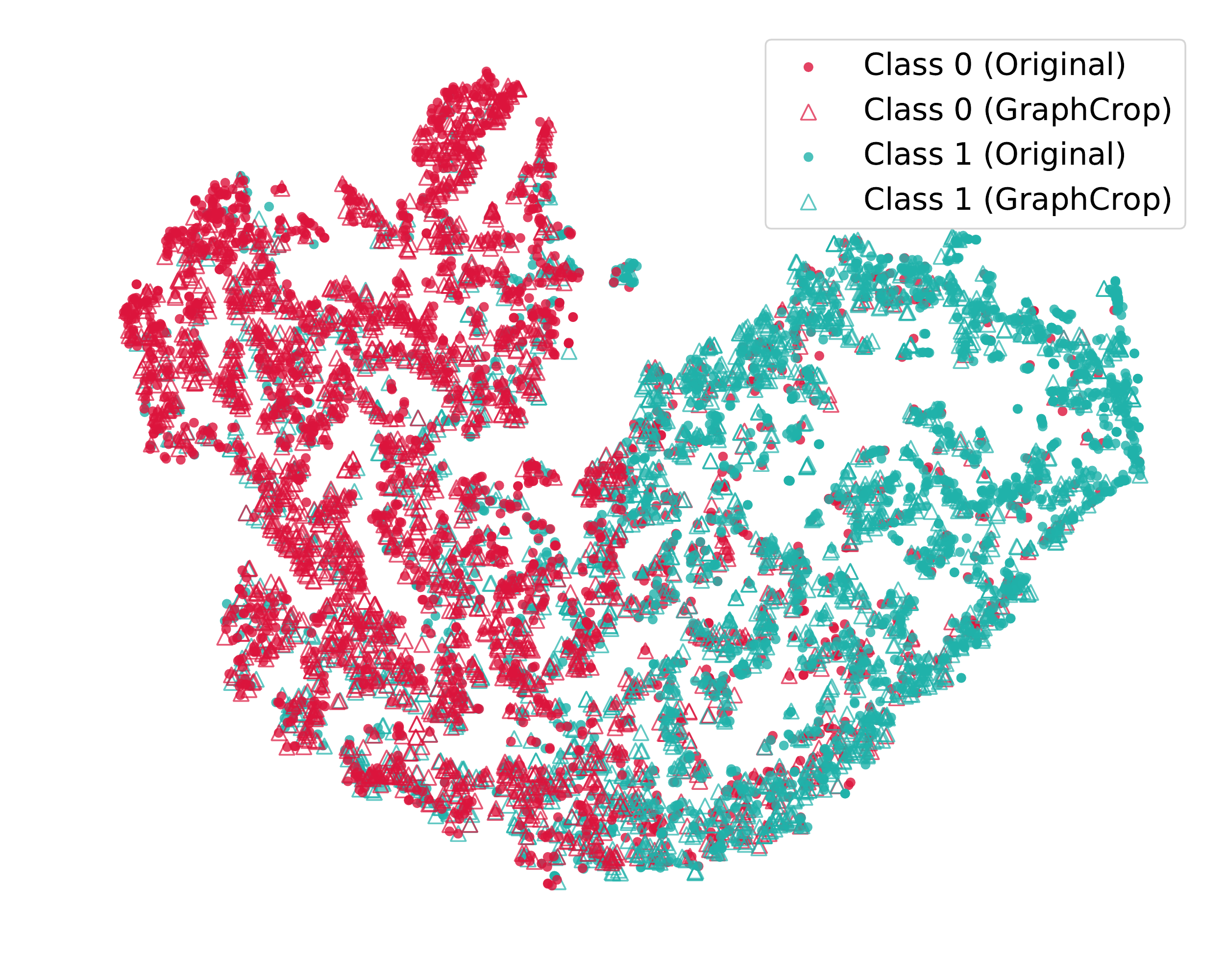}
		\caption{GraphCrop}
	\end{subfigure}
	\caption{\textbf{Augmented graphs (triangles) closely surround original graphs (circles) of the same class (color), suggesting that augmented graphs maintianed their original labels.} We visualize the final-layer graph representations of the original and augmented graphs from the \texttt{NCI109} dataset using t-SNE \cite{van2014accelerating} \label{fig:tsne}}
\end{figure}

In data augmentation, input data is altered while the ground-truth labels are expected to remain unchanged.
If graphs are significantly changed, however, the original class labels may no longer be valid.
We take a visualization approach to examine whether GraphCrop significantly changes the semantic features of augmented graphs.
First, we train a GIN on the \texttt{NCI109} dataset without augmentation.
Then we apply GraphCrop to generate an augmented graph per original graph.
These are fed into the GIN along with the original graphs, and we extract the final-layer representations.
We visualize these graph representations in Fig. \ref{fig:tsne} via t-SNE \cite{van2014accelerating}.
We find that the resulting latent space representations for augmented graphs closely surround those of the original graphs, which suggests that for the most part, graphs augmented with GraphCrop conserve the labels of their original graphs.

\subsection{Ablation Study}
We conduct a number of ablations to analyze GraphCrop.
First, we investigate the effects of the subgraph sampling.
Using our GraphCrop method, we can optionally use the shortest path (SP) or heat diffusion to evaluate connectivity in addition to our default PPR diffusion.
Moreover, we try another two options for subgraph sampling, of which the first one is to uniformly sample all the nodes to retain in the subgraph (UniNode), as visualized in Fig. \ref{fig:uni}.
The second one is DropEdge \cite{rong2019dropedge} designed for node classification, which drops edges uniformly without removing nodes.
We compare the test accuracy of GIN trained with the above-mentioned subgraph sampling techniques in Table \ref{tab:sub_sam}.
Our GraphCrop with different subgraph sampling methods achieves much better performance than UniNode and DropEdge.
The reason is that both UniNode and DropEdge do not guarantee the connectivity between the nodes in the cropped subgraph.
They retain individual nodes or edges rather than the contiguous subgraphs as GraphCrop does.
As a result, the original graph structural characteristics are difficult to maintain in the cropped subgraph.
In contrast, GraphCrop retrieves the subgraphs containing the intact and valid structural context for graph classification.
This encourages GNNs to better utilize the global structural context of a graph, rather than relying on the presence of a few key nodes, which may not be present.
GraphCrop with SP performs worse than GraphCrop with PPR diffusion, since the shortest path distance does not necessarily change when connections (paths) between nodes increase and thus it does not reflect the connectivity effectively.
Empirically, PPR diffusion achieves promising performance.

\begin{table}[tb!]
	\centering
	\caption{Test Accuracy (\%) of graph classification of GIN trained with different subgraph sampling methods.}
	\label{tab:sub_sam}
	\begin{tabular}{@{}l | c c c @{}}
		\toprule
		\textbf{Method}
		& \texttt{D\&D}
		& \texttt{NCI109}
		& \texttt{IMDB-B} \\ \midrule
		UniNode & 75.6 $\pm$ 2.4 & 78.3 $\pm$ 2.1 & 71.5 $\pm$ 3.9\\ 
		DropEdge & 75.5 $\pm$ 2.6 & 78.3 $\pm$ 2.2 & 71.3 $\pm$ 3.7 \\
		GraphCrop (SP) & 76.1 $\pm$ 2.4 & 79.2 $\pm$ 2.3  & 72.5 $\pm$ 3.6\\
		GraphCrop (heat) & 76.4 $\pm$ 2.6 & 79.1 $\pm$ 2.1 & 72.8 $\pm$ 3.8\\
		GraphCrop (PPR) & \textbf{76.6 $\pm$ 2.7} & \textbf{79.5 $\pm$ 2.2} & \textbf{72.9 $\pm$ 3.7}\\
		\bottomrule
	\end{tabular}
\end{table}

\begin{figure}[!tb]
	\centering
	\begin{subfigure}[t]{0.23\textwidth}
		\includegraphics[width=\textwidth]{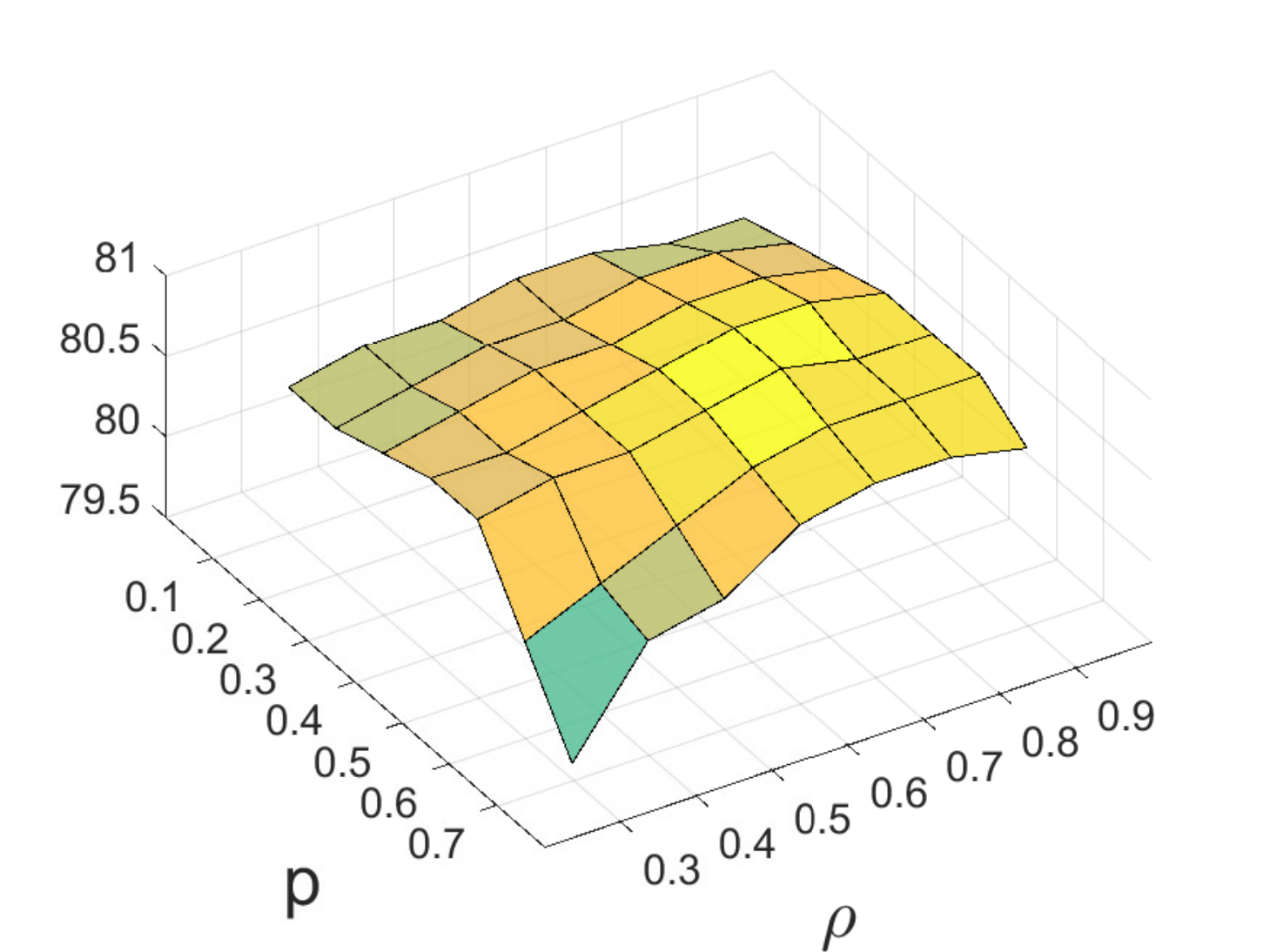}
		\caption{\label{fig:sim2}\texttt{NCI1}}
	\end{subfigure}	\hfill
	\begin{subfigure}[t]{0.23\textwidth}
		\includegraphics[width=\textwidth]{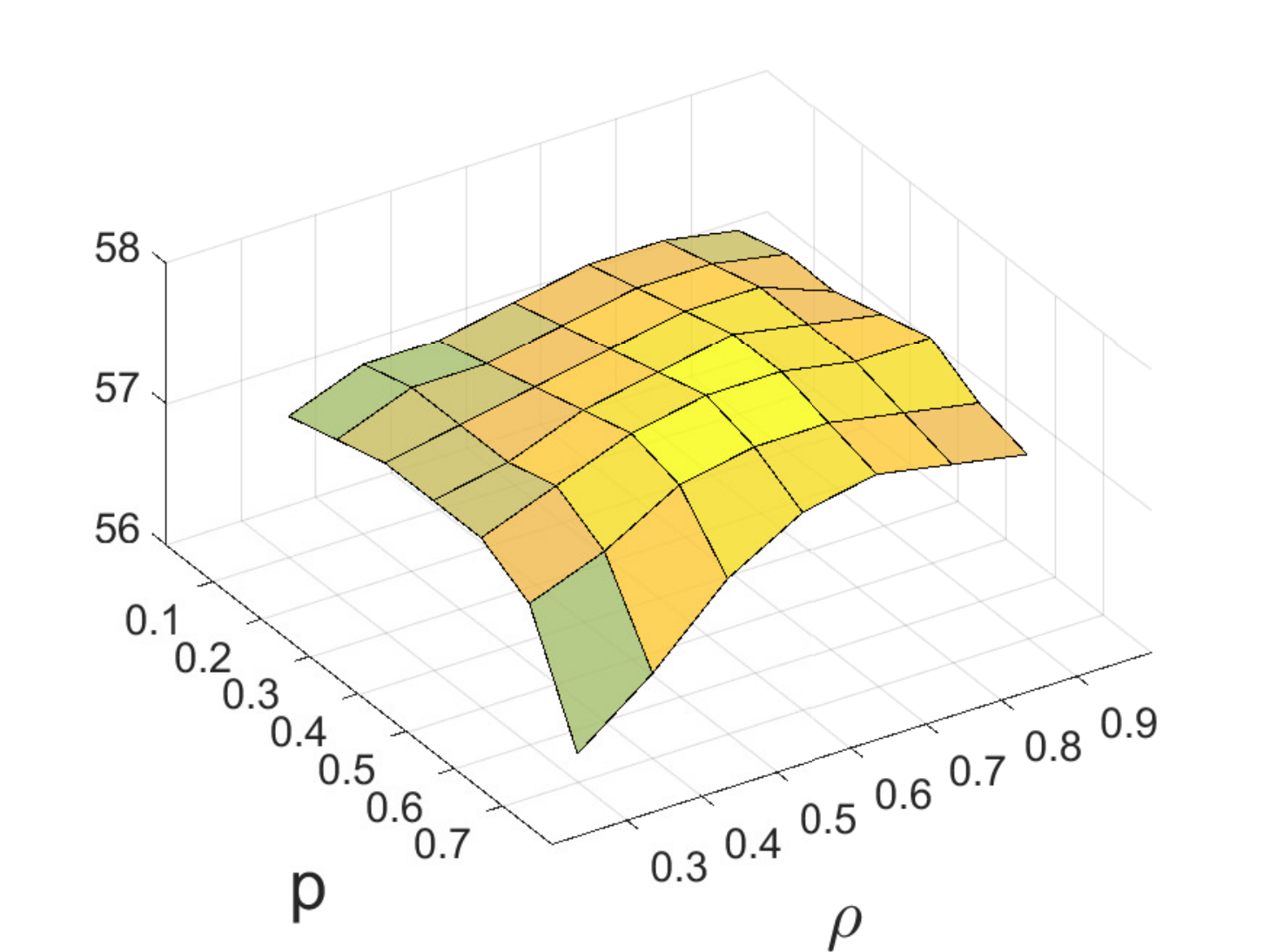}
		\caption{\label{fig:sim3}\texttt{REDDIT-5K}}
	\end{subfigure}
	\caption{The test accuracy (z-axis) of GIN enhanced by GraphCrop with different hyperparameter values $p$ and $\rho$.\label{fig:hyperparameter}}
\end{figure}

Last but not least, we evaluate how sensitive our GraphCrop is to the selection of hyper-parameter values: $p$ to control the probability of our random data augmentation during training and $\rho$ to adjust the subgraph sizes.
As we can see, the performance of GIN with CurGraph is relatively smooth when parameters are within certain ranges.
However, extremely large values of $p$ and severely small values of $\rho$ result in low performances, which should be avoided in practice. 
Moreover, increasing $p$ from 0.1 to 0.5 improves the test accuracy of GIN with GraphCrop, demonstrating that the novel and valid augmented graphs provided by our GraphCrop plays an important role in improving the performance of GNNs.

\section{Conclusion}
In this paper, we propose a new data augmentation approach named GraphCrop (Subgraph Cropping) for graph classification.
GraphCrop uses a node-centric strategy to crop subgraphs that span the contiguous parts of the original graphs, which hold strong inner connectivity.
In principle, it simulates the real-world noise of sub-structure omission to expand the original training set.
By training GNNs with GraphCrop, we regularize GNNs to achieve better generalization virtually for free.
We conduct the experiments on the standard datasets collected from diverse scenarios.
Empirical results show that GraphCrop achieves consistent and substantial improvements for GNN architectures in terms of test accuracy.
In future work, we will explore more data augmentation techniques for graph classification and apply them to other tasks requiring graph-level representations.

\bibliography{GraphClass}

\end{document}